\input{aipcheck}

\documentclass[
    ,final            
  ]
  {aipproc}

\newcommand{\oa}[1]{${\cal O}(a^{#1})$}

\layoutstyle{6x9}

\begin{document}

\title{Effects of dynamical sea-quarks on quark and gluon propagators}


\author{Maria B.~Parappilly}{
  address={Special Research Center for the Subatomic Structure of
Matter (CSSM) and Department of Physics,
University of Adelaide 5005, Australia}
}

\author{Patrick O.\ Bowman}{
  address={Nuclear Theory Center, Indiana University, Bloomington IN 47405,
USA}
}

\author{Urs M.\ Heller}{
  address={American Physical Society, One
Research Road, Box 9000, Ridge, NY 11961-9000, USA}
}

\author{Derek B.\ Leinweber}{
  address={Special Research Center for the Subatomic Structure of
Matter (CSSM) and Department of Physics,
University of Adelaide 5005, Australia}
}

\author{Anthony G.\ Williams}{
  address={Special Research Center for the Subatomic Structure of
Matter (CSSM) and Department of Physics,
University of Adelaide 5005, Australia}
}

\author{J.\ B.\ Zhang}{
  address={Special Research Center for the Subatomic Structure of
Matter (CSSM) and Department of Physics,
University of Adelaide 5005, Australia}
}

\begin{abstract}
We present an unquenched calculation of the quark propagator in Landau gauge 
with 2+1 flavors of dynamical quarks. We study the scaling behavior of 
the quark propagator in full QCD on two lattices with different lattice spacings and
similar physical volume. We use configurations generated with an improved
staggered (``Asqtad'') action by the MILC collaboration.
\end{abstract}

\maketitle


Due to the difficulties of simulating dynamical fermions, 
most of the QCD simulations in the past were done in the quenched 
approximation, that is, ignoring the dynamics of sea quarks. 
In the quenched approximation the determinant of the Dirac operator 
is replaced by a constant.  This would be reasonable if quarks were very
heavy, but as they are not, this approximation results in uncontrolled 
systematic errors that can be as large as 30\% \cite{Davies:2003ik}.

Computing resources now available are powerful enough to treat up, down and 
strange quarks dynamically.  In particular there has been a great deal of 
progress using the staggered formalism for lattice fermions.  We have 
calculated the gluon and quark propagator in Landau gauge using configurations
generated by the MILC collaboration~\cite{Bernard:2001av,Aubin:2004wf} 
available from the Gauge Connection~\url{http://www.qcd-dmz.nersc.gov}.

  
The MILC configurations were generated with the \oa{2} one-loop
Symanzik-improved L\"{u}scher--Weisz gauge action.
The dynamical configurations use the Asqtad quark action,
an \oa{2} Symanzik-improved staggered fermion action.
First results for the gluon propagator and quark propagator in full QCD were
published in Ref.~\cite{Bowman:2004jm} and Ref.~\cite{Bowman:2005vx} 
respectively.  We here extend those results by using a finer lattice.
As well as being interesting in themselves, the study of the propagators
is proving to be a fruitful area of interaction between lattice gauge theory
and Dyson-Schwinger equations.  See, for example, 
Refs.~\cite{Alkofer:2003jj,Bhagwat:2003vw}.


\subsection{Effects of dynamical sea-quarks}

First we discuss the gluon dressing function, $q^2D(q)$, where 
$D(q) = \langle A(q) A(-q) \rangle$, is the gluon propagator.  The addition of 
dynamical quarks to the gauge fields produces a clearly 
visible effect in the dressing function in the region of the infrared hump.
Unquenching results in a reduction of around $30$\% at 1 GeV.
The qualitative features of the propagator -- enhancement of the
intermediate infrared momenta followed by suppression in the deep
infrared -- are, however, unchanged.  Spectral positivity is violated in 
full QCD just as in the quenched theory, something that will be discussed in
more detail in an upcoming publication.

\begin{table}
\begin{tabular}{lrrrr}
\hline
    \tablehead{1}{r}{b}{Dimensions}
  & \tablehead{1}{r}{b}{$\beta$ }
  & \tablehead{1}{r}{b}{ $a$}
  & \tablehead{1}{r}{b}{Bare Quark Mass}   
  & \tablehead{1}{r}{b}{\#Config} \\  
\hline
  $28^3\times 96$ &   7.09  & 0.090 fm  &$14.0$ MeV, $67.8$ MeV  & 108 \\
  $28^3\times 96$ &   7.11  & 0.090 fm  &$27.1$ MeV, $67.8$ MeV  & 110 \\
\hline 
  $20^3\times 64$ &   6.76  & 0.125 fm  &$15.7$ MeV, $78.9$ MeV  & 203 \\ 
  $20^3\times 64$ &   6.79  & 0.125 fm  &$31.5$ MeV, $78.9$ MeV  & 249 \\ 
  $20^3\times 64$ &   6.81  & 0.125 fm  &$47.3$ MeV, $78.9$ MeV  & 268 \\
  $20^3\times 64$ &   6.83  & 0.125 fm  &$63.1$ MeV, $78.9$ MeV  & 318\\ 
\hline
\end{tabular}
\caption{Lattice parameters used in this study. The
dynamical configurations each have two degenerate light quarks
(up/down) and a heavier quark (strange).}
\label{tab:a}
\end{table}

\begin{figure}[t]
\centering\includegraphics[height=.6\textheight,angle=90]{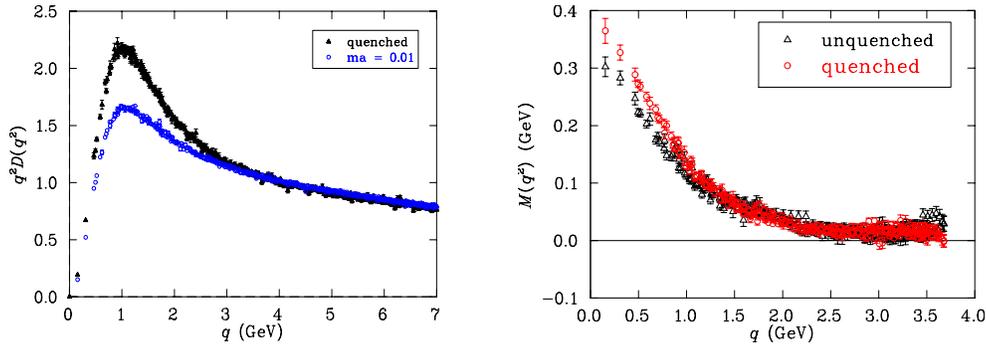}

\caption{ The gluon dressing function in Landau gauge is left.  Full triangles
correspond to the quenched calculation, while open circles correspond
to 2+1 flavor QCD.  As the lattice spacing and volume are the same,
the difference between the two results is entirely due to the presence
of quark loops. Right is the comparison of the unquenched (full QCD) and quenched quark
propagator for non-zero quark mass.  The mass function for the
unquenched dynamical-fermion propagator has been interpolated so that
it agrees with the quenched mass function for $ma = 0.01$ at the renormalization point, 
$q$ = 3 GeV.
For the unquenched propagator this corresponds to a bare quark mass of
$ma = 0.0087$. }
\label{fullgq}
\end{figure}

On the right-hand side of Fig.~\ref{fullgq} we compare both quenched and dynamical data 
for the quark mass function. For the comparison, we select a bare quark mass for the quenched
case ($ma =0.01$) and interpolate the dynamical mass function so that it agrees with the quenched 
result at the renormalization point, $q=3$ GeV.   
The quark propagator is not strongly altered by the presence of
sea quarks.  The dynamical mass generation is somewhat supressed, the quark
mass function at zero four-momentum being reduced by about $20$\% in the
chiral limit.  For a given bare quark mass, the running mass is larger in 
full QCD than in quenched QCD.  The wavefunction
renormalization function, $Z$, is insensitive to changes in the bare quark 
mass for the cases studied here.

\begin{figure}[t]
\centering\includegraphics[height=.35\textheight,angle=90]{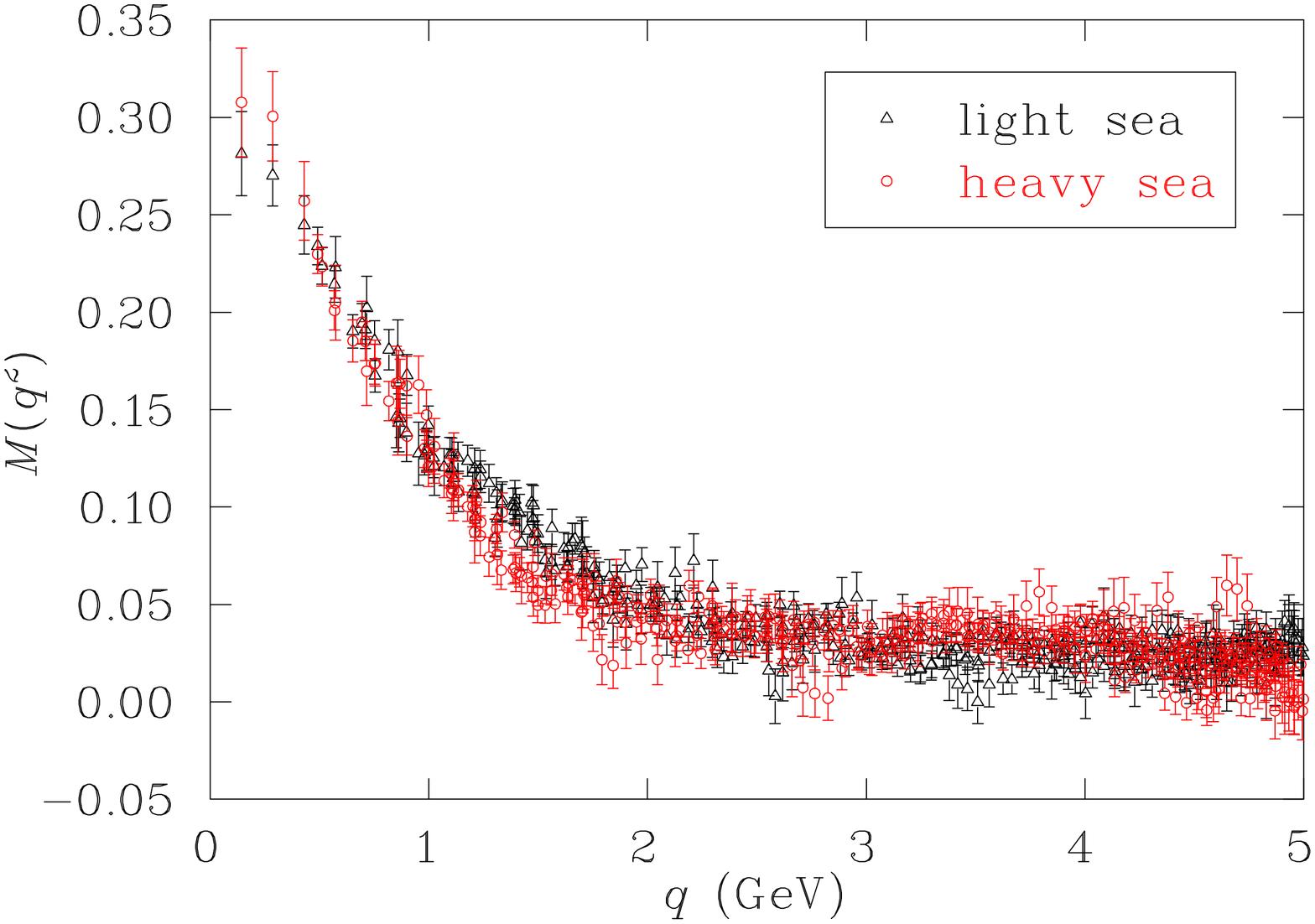}
\centering\includegraphics[height=.35\textheight,angle=90]{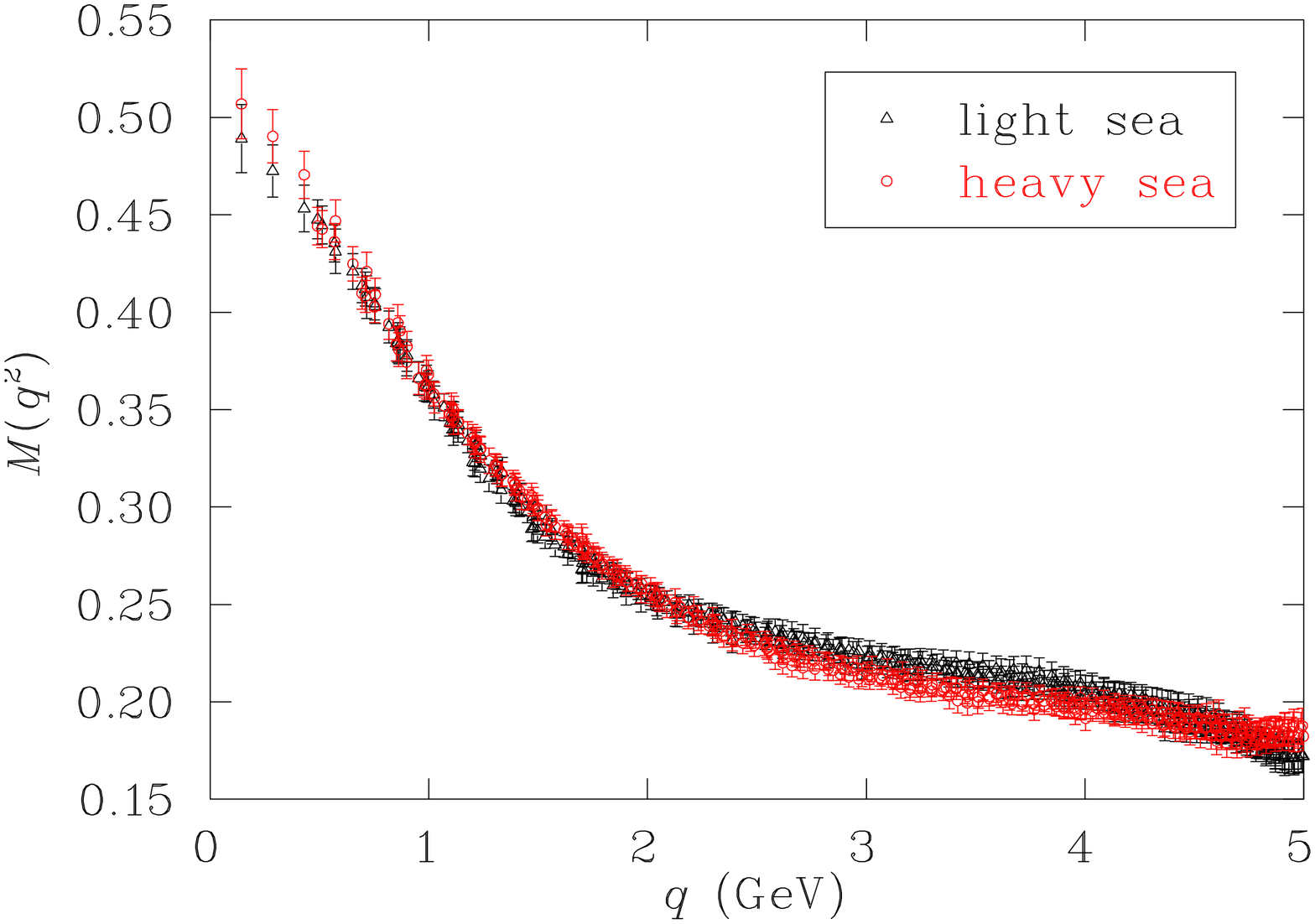}
\caption{The unquenched quark mass function for the two different values of the
light sea quark mass on the fine lattice ($14.0$ MeV and $27.1$ MeV). The valence quark masses are 
$m = 14.0$ MeV (left) and  $m = 135.6.$ MeV (right), the lightest and heaviest in our current sample 
respectively.}
\label{seabiglat}
\end{figure}

In Fig.~\ref{seabiglat} the valence quark mass is held fixed while the
sea quark mass changes.  Clearly the dependence over this small range of sea 
quark masses is weak.  Unfortunately we only have two dynamical sets to 
compare, and for the lightest valence quark the data are rather noisy.
We have studied the scaling behavior of quark propagator by  
working on two lattices with different lattice spacing but similar 
physical volume.  We compared the wave-function renormalization function 
$Z(q^2)$ and mass function $M(q^2)$ for two lattices in
Ref.~\cite{Parappilly:2005ei}.  The quark propagators are in 
excellent agreement,showing no observable dependence on the lattice spacing.

These results reflect the fact that quenched QCD is in some sense, ``maximally
nonabelian.''  The quark loops compete with the self-interacting gauge field, 
screening the color charge and weakening -- but by no means overcoming -- 
confinement and chiral symmetry breaking.  Furthermore, we find good scaling 
behavior for Asqtad fermions at a lattice spacing of $a=0.125$ fm.


 
{\bf Acknowledgments} : We thank the South Australian Partnership for Advanced Computing (SAPAC)
for generous grants of supercomputer time which have enabled this
project.




\bibliographystyle{aipproc}   

\bibliography{panic05}

\IfFileExists{\jobname.bbl}{}
 {\typeout{}
  \typeout{******************************************}
 \typeout{** Please run "bibtex \jobname" to optain}
  \typeout{** the bibliography and then re-run LaTeX}
  \typeout{** twice to fix the references!}
  \typeout{******************************************}
  \typeout{}
 }

\end{document}